\documentstyle[epsf,floats,twocolumn,prb,aps]{revtex}

\begin{document}
\draft

\title{The oxidation state at tunnel junction interfaces}

\author{X.~Batlle \cite{XBatlle}, B.J.~Hattink, A.~Labarta}
\address{Dept. F\'{\i}sica Fonamental, Univ. Barcelona, Av. Diagonal
  647, 08028-Barcelona, Catalonia, Spain}

\author{B.J.~J\"onsson-\AA kerman, R.~Escudero \cite{REscudero},
  I.K.~Schuller}
\address{University of California --- San Diego, Physics
  Department-0319, La Jolla, CA 92093-0319, USA}

\date{\today}
\maketitle

\begin{abstract}
  We demonstrate that at the usual $10^{-7}$ torr range of base
  pressures in the sputtering chamber, X-ray photoelectron
  spectroscopy shows the existence of a thin AlO$_x$ layer at the
  Nb/Al interface in both Nb/Al-AlO$_x$/Pb tunnel junctions and Nb/Al
  bilayers. This is due to the time elapsed between the deposition of
  the Nb and Al bottom layers, even at times as short as 100~s. We
  also give some direct evidence of the oxidation of the top Pb
  electrode on the Nb electrode surface. Such oxidation probably
  occurs at the pinholes of the intermediate Al-AlO$_x$ layer of the
  junctions, as a consequence of the oxidation state at the Nb/Al
  interface. We therefore suggest that both the base pressure and the
  time lapse between layer depositions should be carefully controlled
  in magnetic tunnel junctions.
\end{abstract}



Ferromagnet/insulator/ferrromagnet (FM/I/FM) magnetic tunnel junctions
(MTJs) exhibiting large magnetoresistance (MR)\cite{Mod95} have lately
attracted much interest due to their potential
applications\cite{Dau97,Gal97}. The performance of the junctions is
strongly dependent on the oxidation of the FM electrodes at the FM/I
interfaces, as well as on the oxidation state of the barrier, which
has to be homogenous and complete. The use of thinner and thinner
barriers has reopened the question of how to rule out the presence of
pinholes. Rowell and others developed a set of criteria to ascertain
that tunneling is the dominant mechanism in junctions with at least
one superconducting (S) electrode\cite{Bur69}. Three of these criteria
still apply in FM/I/FM structures: (i) an exponential insulator
thickness dependence of the conductance, $G$; (ii) a parabolic voltage
dependence of $G$ that can be fitted to the theoretical
models\cite{Bri70,Sim63}; and (iii) a weak insulating-like dependence
$G(T)$.

For the first criterion, it has been shown\cite{Rab01} that pinholes
may mimic the exponential thickness dependence of the tunneling
resistance. For the second one, some of us demonstrated\cite{Jon00}
that S/I/FM junctions that displayed parabolic $G(V)$ curves in the
normal state, showed, at low temperatures and depending on the
oxidation procedure, either tunneling or pinhole conduction (Andreev
reflection\cite{Blo82}). The recent observation of very large MR in
Co-Co and Ni-Ni wire nanocontacts\cite{Tat99} also suggests the
pinhole contribution to TMR in MTJs. Given the ratio of the conduction
between metallic contacts and junctions\cite{Pra91}, pinhole regions
of one part in $10^6$ must be ruled out to ensure no pinhole
conduction in parallel with tunneling.  For the third criterion, some
results suggest that pinholes yield a metallic-like temperature
dependence of the junction resistance\cite{Jon00,Rue01}. Therefore,
out of the three Rowell criteria, only one, the insulating-like
$G(T)$, seems to be reliable.

X-ray photoelectron spectroscopy (XPS)\cite{Bri83} is an excellent
technique for the analysis of MTJs\cite{Qiu00} since it is sensitive
to the chemical species as well as to their bonding state.  In this
letter, we show that at the usual $10^{-7}$ torr range of base
pressures in the sputtering chamber, XPS indicates that there exists a
thin oxygen layer mostly adsorbed on the surface of the bottom
electrode that yields the oxidation of the first impinging atoms of
the intermediate layer. We also demostrate that this is due to the
time elapsed between the deposition of the layers. Therefore, the
oxidation state at the interfaces strongly affects the tunneling
process by modifying the pinhole conductivity and the interface
chemistry and roughness.

The junctions were prepared as follows\cite{Jon00}: a dc sputtered
superconducting Nb(80~nm)/Al(10~nm) bilayer bottom electrode was
oxidized in air 10 minutes and a Pb top electrode (200~nm) was
deposited in a separate thermal evaporation unit, leading to the
Nb/Al-AlO$_x$/Pb junction (Pb sample). The resulting oxide thickness
was typically 1-2~nm and the remaining 8~nm of metallic Al were
superconducting for proximity to the Nb\cite{Zeh99}. For a second
sample, the bottom electrode was oxidized for 18 hours in air (barrier
thickness of 2-3~nm) and a Ni (20~nm) layer was dc sputtered on top,
leading to the Nb/Al-AlO$_x$/Ni junction (Ni sample). The junction
area for both samples was $1 \times 0.3$~mm$^2$. These two samples are
representative of the variety of samples studied\cite{Jon00}. Standard
ac (1 kHz) differential conductance $G=dI/dV$ measurements as a
function of dc bias were carried out from 4.2 to 300~K using a
balanced bridge. XPS spectra (Al K$_\alpha$; base pressure $10^{-9}$
torr) were recorded for exactly the same samples as in $G(V)$. The
distribution of elements across the junction was studied by performing
a low-energy sputtering process (4~keV, incident at 45$^\circ$;
typical etching rate 6-10~nm/minute) for a short time (6-18~s) and
recording the spectrum after each step.  This sputtering process,
which may lead to a certain intermixing, together with the fact that
the XPS signal averages the out-coming electrons from a region of
about 5-10~nm in depth, precludes the observation of sharp interfaces.

\begin{figure}[t]
\centering
\leavevmode
\epsfxsize=6cm
\epsfbox{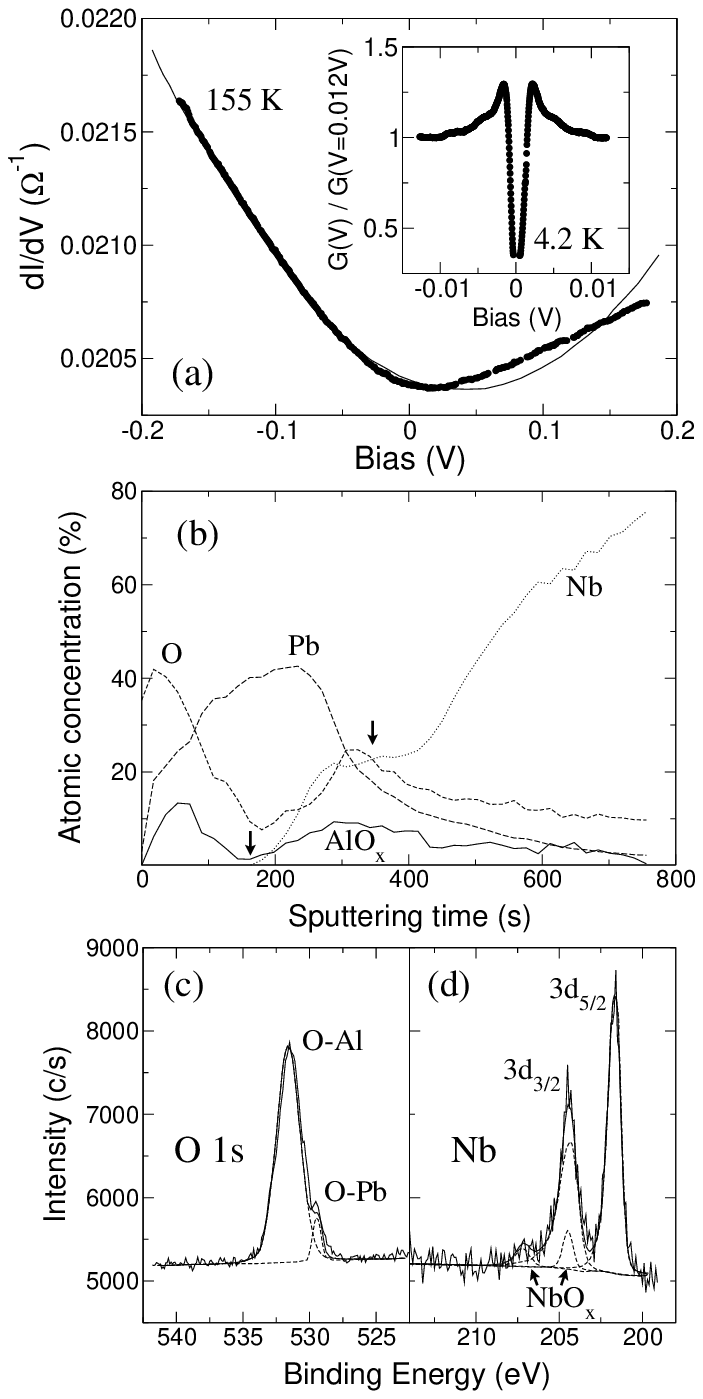}
\vspace{0.4cm}
\caption{Nb/Al-AlO$_x$/Pb junction: (a) $G(V)$ curve at 155~K, together
  with a fit to the BDR model. Inset. $G(V)$ curve at 4.2~K, showing
  the superconducting gap (tunnel conduction). (b) Atomic
  concentration obtained from the XPS intensities for Pb, O, AlO$_x$
  and Nb, as a function of the sputtering time. (c) Example of the XPS
  spectra for the O $1s$ core level at the Nb/Al interface (sputtering
  time: 324~s), showing the O-Al and O-Pb (3\%) contributions. (d) Nd
  $3d$ core level (sputtering time: 216~s), showing the NbO
  contribution.}
\label{fig:G_XPS}
\end{figure}

For the Pb junction, $G(V)$ at 155~K suggests tunnel conduction
(Fig.~\ref{fig:G_XPS}(a)), which is confirmed by the signature of the
superconducting gap at 4.2~K (inset of Fig.~\ref{fig:G_XPS}(a)). The
fit of $G(V)$ to the Brinkman-Dynes-Rowell (BDR) model\cite{Bri70}
yields a barrier thickness of 1.1~nm and heights of 3.6~eV and 2.4~eV
on the Pb and Nb/Al sides, respectively, while the fit to the Simmons'
model\cite{Sim63} yields 1.2~nm and 2.9~eV. For the Ni sample, $G$ is
high and flat, indicating big metal-to-metal shorts. This suggests
that Pb mostly adds to the AlO$_x$ barrier without damaging it and
that the latter is free of metal-to-metal pinholes.  However, Ni
damages the barrier, in agreement with previous studies of the
influence of the top electrode on the junction properties: both the
junction resistance\cite{Han62} and the barrier height\cite{Nin88}
increase with the ionic radius (from Ni to Pb) since the larger the
latter is, the less the element can penetrate in the barrier,
resulting in less effective micro-shorts.

Fig.~\ref{fig:G_XPS}(b) shows the atomic concentration of Nb, Pb,
AlO$_x$ and O obtained from the XPS intensities\cite{Bri83} for the Pb
sample, as a function of the sputtering time (sputtering steps of
18~s). The AlO$_x$ concentration was obtained by fitting the intensity
of the Al $2p$ core level to both a metallic and oxide
contribution\cite{Bri83}. As soon as Nb is detected, the O signal
increases, first reaching a maximum and later decreasing as the
sputtering process approaches the Nb layer.  The AlO$_x$ concentration
perfectly follows that of O (Fig.~\ref{fig:G_XPS}(b)), the first
maximum corresponding to the insulating barrier, the second one (ca.
300~s) suggesting that there is a thin oxygen layer on top of Nb that
oxidizes the first Al atoms arriving at the bottom electrode. At short
sputtering times, the oxidation of both the free surface of the sample
(O-Pb $1s$ at 528.3-529.4~eV) and the barrier (O-Al $1s$ at
531-531.5~eV) are observed. Besides, the O-Al $1s$ peaks at the Nb/Al
interface are clearly asymmetric at low energies, and a small peak at
about 529.3~eV is observed (Fig.~\ref{fig:G_XPS}(c)), suggesting an
O-Pb contribution of about 3\%. This indicates that some PbO$_x$ is
present on the Nb surface, which is probably due both to the existence
of the oxygen layer at the Nb/Al interface and to the consequent Pb
oxidation at the pinholes of the Al-AlO$_x$ layer. This is in
agreement with some electron microscopy studies showing a pinhole
surface area up to about 2\% in AlO$_x$ barriers\cite{Ozk00}, even if
in the present Pb junction, the former Al layer was about 10~nm in
thickness. From the Nb $3d$ core level (Fig.~\ref{fig:G_XPS}(d)), some
NbO (Nb-O $3d_{5/2}$ at 204.6~eV) on the surface of the Nb layer is
inferred, although the calculated O-Nb contribution is at least one
order of magnitude smaller than the experimental O-Al one, so that the
former (530.0-530.4~eV) is not detected since it overlaps to the
latter. The Pb $4f$ core level (not shown) at the Nb/Al interface also
yields a small amount of PbO and/or Pb$_3$O$_4$, while the O-Pb/Pb-O
ratio is about 1.5. The Ni sample also follows this framework: both
the O-Al $1s$ and Al-O $2p$ signals increase on the surface of Nb.

Given the base pressures in the sputtering chamber ($2.6\times
10^{-7}$ and $1.3\times 10^{-7}$ torr for the Pb and Ni samples,
respectively), the thin oxygen layer at the Nb/Al interface is related
to the time that the Nb film is kept in the chamber before the Al
deposition proceeds (45 minutes and 5 hours for the Pb and Ni
junctions, respectively). Three Nb(100~nm)/Al(20~nm) bilayers were
prepared to support this suggestion: sample Nb/Al-1, for which the Nb
layer was 100~s in chamber (base pressure of $1.4\times 10^{-7}$
torr); sample Nb/Al-2, for which the Nb layer was 18.5 hours in the
chamber ($1.2\times 10^{-7}$ torr); and sample Nb/Al-3, for which the
Nb layer was 27 hours in air (former base pressure of $1.4\times
10^{-7}$ torr). The role of the standard Gibbs energy of formation of
the metal oxides ($\Delta_f G_0$ at 298.15~K)\cite{CRC} was studied by
preparing two Nb(100~nm)/Pb(200~nm) bilayers with increasing transfer
time in air from the sputtering chamber to the Pb evaporation unit;
sample Nb/Pb-1 was 7 minutes in air, while sample Nb/Pb-2, was 24
hours.

\begin{figure}[t]
\centering
\leavevmode
\epsfxsize=6cm
\epsfbox{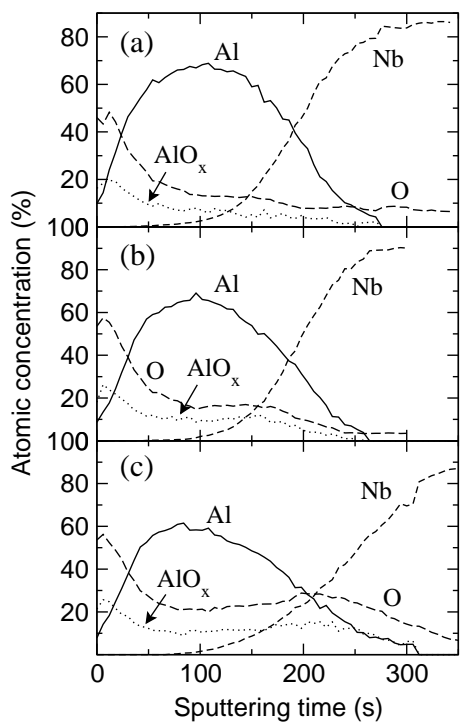}
\vspace{0.3cm}
\caption{(a) Atomic concentration obtained from the XPS intensities for
  Al, O, AlO$_x$ and Nb, as a function of the sputtering time, for
  sample Nb/Al-1 (Nb layer kept 100~s in the sputtering chamber; base
  pressure of $1.4\times 10^{-7}$ torr). (b) Same data for sample
  Nb/Al-2 (18.5 hours, $1.2\times 10^{-7}$ torr). (c) Same data for
  sample Nb/Al-3 (27 hours in air).}
\label{fig:XPS_time}
\end{figure}

The atomic concentration for Nb, metallic Al, O and AlO$_x$ for the
Nb/Al bilayers is shown in Fig.~\ref{fig:XPS_time} as a function of
the sputtering time (sputtering step of 6~s). The AlO$_x$
concentration at the Nb/Al interface clearly increases with increasing
time between the Nb and Al depositions. Even for sample Nb/Al-1
(Fig.~\ref{fig:XPS_time}(a)), there is change in the slope of both the
concentrations of AlO$_x$ and O as the Nb layer is detected.  The O
$1s$ core level yields mostly O-Al bonds and, although there might be
some NbO on the surface of the Nb, the calculated O-Nb contribution is
at least about one order of magnitude smaller than the O-Al one for
all samples, in agreement with $\Delta_f G_0$(Al$_2$O$_3$)=
-1582.3~kJ/mol and $\Delta_f G_0$(NbO)= -378.6~kJ/mol\cite{CRC}. A
similar picture is drawn for the Nb/Pb samples (not shown), the main
differences being that, while at the free surface of the samples the
oxidation state corresponds to O-Pb, in agreement with the Pb-O
contribution to the Pb $4f$ core level, the O $1s$ increase at the
Nb/Pb interface takes place at 530.0-530.4~eV, suggesting that this is
the core level energy for the O-Nb $1s$ bonds. Both the Nb-O peak
position and the O-Nb/Nb-O ratio at the Nb/Pb interfaces, suggest
NbO$_2$.  Besides, the calculated O-Pb contribution is about two
orders of magnitude smaller than the O-Nb one, in agreement with
$\Delta_f G_0$(NbO$_2$)= -740.5~kJ/mol and $\Delta_f G_0$(PbO)= -
187.9~kJ/mol\cite{CRC}.

In conclusion, we have shown that when preparing tunnel junctions at
the usual $10^{-7}$ torr base pressures, the oxidation state at the
interfaces depends on the time elapsed between the deposition of the
layers, even at times as short as 100~s, which is most relevant for
the performance of the junctions. We therefore suggest that both
parameters should be carefully controlled. We have also given some
direct evidence of the oxidation of the top electrode on the surface
of the bottom one, probably at the pinholes of the intermediate layer
---even if the latter is about 10~nm in thickness--- and due to the
oxidation state at the interfaces. As some junctions studied show
tunnel conduction, the oxidation state at the interfaces may be
relevant to the dominant conduction mechanism. Consequently, the
oxidation of the top FM layer at the pinholes of the insulating
barrier in MTJs may well be influenced by the synthesis conditions so
that the role of the latter in the quality of the junctions should be
taken into account, the relative metal oxide formation energy of the
selected elements being crucial.

Financial support of the Spanish CICYT (MAT2000-0858) and the
Generalitat de Catalunya (ACI2000-04 and 2000SGR-025) are gratefully
recognized. IKS acknowledges US DARPA and ONR.

\end{document}